\documentclass[aps,a4paper,showpacs,superscriptaddress,twocolumn]{revtex4-1}
\usepackage{tensor}
\usepackage{graphicx}
\usepackage{amsmath}
\usepackage{amssymb}
\usepackage{enumerate}
\usepackage{subfigure}
\usepackage{tabularx}
\usepackage[colorlinks=true, pdfstartview=FitV, linkcolor=blue, citecolor=red, urlcolor=black, breaklinks=true]{hyperref}
\usepackage{orcidlink}
\newcommand{\be}{\begin{equation}}
\newcommand{\ee}{\end{equation}}
\newcommand{\ben}{\begin{eqnarray}}
\newcommand{\een}{\end{eqnarray}}
\newcommand{\bes}{\begin{subequations}}
\newcommand{\ees}{\end{subequations}}
\def\bal#1\eal{\begin{align}#1\end{align}}

\newcommand{\sech}{{\rm sech}}

\begin{document}
\title{Radially symmetric scalar field solutions in the presence of cuscuton term}
\author{D. Bazeia\,\orcidlink{0000-0003-1335-3705}}
\email{bazeia@fisica.ufpb.br}
\affiliation{Departamento de F\'\i sica, Universidade Federal da Para\'\i ba, 58051-970 Jo\~ao Pessoa, PB, Brazil}
\author{M.A. Marques\,\orcidlink{0000-0001-7022-5502}}
\email{marques@cbiotec.ufpb.br}
\affiliation{Departamento de Biotecnologia, Universidade Federal da Para\'\i ba, 58051-900 Jo\~ao Pessoa, PB, Brazil}
\author{R. Menezes\,\orcidlink{0000-0002-9586-4308}}
     \email[]{rmenezes@dcx.ufpb.br}\affiliation{Departamento de Ci\^encias Exatas, Universidade Federal
da Para\'{\i}ba, 58297-000 Rio Tinto, PB, Brazil}
\author{M. Paganelly\,\orcidlink{0000-0002-0342-2079}}
\email[]{matheuspaganelly@gmail.com}
\affiliation{Departamento de F\'{\i}sica, Universidade Federal de Campina Grande, 58429-900, Campina Grande, PB, Brazil}
\begin{abstract}
In this work, we investigate radially symmetric solutions in arbitrary dimensions in scalar field models in the presence of the cuscuton term. We introduce a first-order formalism compatible with the equation of motion which supports field configurations engendering minimum energy and show that the cuscuton term does not induce instabilities in the solutions. To illustrate the general results, we study two distinct classes of models and present analytical solutions and the corresponding energy densities.
\end{abstract}
\maketitle 

\section{Introduction}
Topological structures are of interest in many branches of Physics since these configurations can arise in scenarios in which spontaneous symmetry breaking is present. The cosmic evolution of the Universe, initially dense and warm, experienced an expansion that may have led to phase transitions which created favorable conditions for the emergence of topological structures \cite{kibble1,kibble2,zurek,VS,MS}. 

Some of the most known topological objects are kinks, vortices, and monopoles, which appear in one, two, and three spatial dimensions, respectively. The equations of motion that govern them are nonlinear differential equations of second order. An important tool in the investigation of these structures is the so-called BPS formalism, introduced by Bogomol’nyi \cite{Bogomolny:1975de}, Prasad and Sommerfield \cite{Prasad:1975kr}, which is based on energy minimization and leads to first-order equations that support stable topological solutions. 
 
One of the simplest topological structures are kinks, which arise under the action of a single real scalar field \cite{Baz,Vacha,Shnir}. In Minkowski spacetime, it is known that Lorentz-invariant scalar field models can only support stable solutions in one spatial dimension; see Derrick and Hobart results in Refs.~\cite{derrick,hobart}. To circumvent the Derrick-Hobart scaling arguments, it was shown in Ref.~\cite{prlbazeia} that one may include an explicit dependence on the radial coordinate in the potential term in the Lagrangian density. This idea was further extended to global monopole-like solutions in Ref.~\cite{global2}. Another approach, introduced in \cite{global1}, consists of taking maximally symmetric multiplet of scalar fields in models with noncanonical kinetic terms in the Lagrangian density.

In the direction of models with non-canonical kinetic terms, one finds the cuscuton term \cite{Afshordi:2006ad,Afshordi:2007yx}, which has gained much attention recently; see, for instance, Refs.~\cite{gomes,iyonaga,kim,bartolo,mylova}. In particular, in \cite{gomes}, the authors perform the Hamiltonian analysis of the cuscuton theory, showing the presence of a singular behavior in the homogeneous limit with local degrees of freedom. Also, Ref.~\cite{iyonaga} introduces a general class of theories with the same property as the cuscuton, which was called extended cuscuton. In \cite{kim}, the authors extend  the cuscuton theory for a bouncing Universe, wherein they investigate an alternative to inflation for generating near scale-invariant scalar perturbations from the cuscuton bounce. In \cite{bartolo}, the investigation focuses on the study of the effects of the cuscuton in the context of inflation. In addition, Ref.~\cite{mylova} examines the construction of an effective geometric formulation for the cuscuton field theory, incorporating curvature-based modifications into the original surface and volume terms in the action.

The standard cuscuton model engenders a field with infinite sound speed, which does not contribute to temporal derivative terms in the equation of motion. Recent works dealing with topological structures have investigated how the presence of the cuscuton modifies the behavior of the solutions \cite{cuscutonkink,Lima:2021ekz,Bazeia:2022sgb,Andrade:2023jvf}. In particular, in Ref.~\cite{cuscutonkink}, it was shown that the cuscuton term does not appear in the equation of motion and, therefore, preserves the standard (cuscuton-free) kink solution, which remains stable under small fluctuations. In the braneworld scenario with a single extra dimension of infinite extent, the cuscuton term appears in the contribution of the geometry in the equation of motion associated to the scalar field, modifying the brane profile and its stability potential, as explicitly shown in \cite{Bazeia:2022sgb}. 

In this paper, we investigate another novel issue, concerning the possibility to obtain stable radially symmetric scalar field solutions in models which include the cuscuton term in arbitrary dimensions. The work is organized as follows: in Sec.~\ref{section2} the properties of the model are investigated, such as the equation of motion and the energy density. Also, a first-order framework based on energy minimization is developed. In Sec.~\ref{stabilitysec}, the stability of the solutions under small fluctuations is studied. There, it is shown that the fluctuations are governed by a Sturm-Liouville eigenvalue equation. In Secs.~\ref{fm} and \ref{sm} we illustrate our procedure with two distinct models in the scenarios of arbitrary spatial dimensions. The first model leads to solutions that can be fully mapped on kinks, differing from the results known so far. The second case constructs a model that fits the solution with power-law tails governed by an auxiliary function that depends on the spatial dimension $D$. We end this work in Sec.~\ref{finalremarks} presenting some comments and perspectives for future research.

\section{Generalities}\label{section2}
We start the investigation with the action of a single real scalar field in $(D,1)$ flat spacetime dimensions 
\be
    S=\int dt\, d^Dx \left(X-\frac{2f(\phi)X}{\sqrt{|2X|}}-V(\phi)\right),
\ee
where
\be
X =\frac{1}{2}\partial_{\mu}\phi\partial^{\mu}\phi.
\ee
Here, we use the metric tensor with mostly minus components, $\eta_{\mu\nu}=\text{diag}(+1,-1,-1,-1,...)$. The function $f(\phi)$ controls the strength of the cuscuton term and $V(\phi)$ is the potential. By varying the action with respect to the field, we get the following equation of motion
\be
\label{emtd}
    \bigg[ \bigg(1-\frac{f(\phi)}{\sqrt{|2X|}}\bigg)\eta^{\mu\nu}+\frac{f\partial^{\mu}\phi\partial^{\nu}\phi}{2X\sqrt{|2X|}}\bigg]\partial_{\mu}\partial_{\nu}\phi+V_{\phi}=0.
\ee
As it is known from Derrick-Hobart arguments \cite{derrick,hobart}, scalar field models only support stable static solutions in a single spatial dimension. In Ref.~\cite{prlbazeia}, it was presented a procedure to circumvent this issue, in which the spherical symmetry was assumed. We follow these lines with static configurations, $\phi=\phi(r)$, to write
\begin{align}
\label{em}
    \frac{1}{r^{D-1}}(r^{D-1}\phi ')'-\frac{(D-1)}{r}\frac{f\,\phi'}{|\phi'|}=V_{\phi}.
\end{align}
The prime represents derivative with respect to the radial coordinate, $\phi'=d\phi/dr$. In this situation, $X=-{\phi^\prime}^2/2$. Note that the second term disappears for $D=1$, recovering the model studied in \cite{cuscutonkink}. For general $D$, we impose that the function $f(\phi)$ vanishes at the points in which $\phi'$ is null in order to avoid discontinuity in its respective term. The above equation shows that, even in higher dimensions, the cuscuton term does not contribute with second-order derivatives in the equation of motion. 

The energy density is calculated standardly; it has the form
\be\label{rhodens}
    \rho=\frac{\phi'^{2}}{2}-f|\phi'|+V.
\ee
For monotonically increasing solutions, we have $\phi'\geq0$. In this case, one can introduce an auxiliary function $W(\phi)$ with $W_\phi\geq0$ where the solution exists, to write the above expression as
\be\begin{aligned}
    \rho &=\frac{1}{2}\bigg(\phi'-f(\phi)-\frac{W_{\phi}}{r^{D-1}}\bigg)^{2}+V
    \\
    &-\frac{1}{2}\bigg(\frac{W_{\phi}}{r^{D-1}}+f(\phi)\bigg)^2+\frac{W'}{r^{D-1}}.
    \end{aligned}
\ee
If the potential has the form
\be
\label{poten}
    V=\frac{1}{2}\bigg(\frac{W_{\phi}}{r^{D-1}}+f(\phi)\bigg)^2,
\ee
one can see that the energy is bounded,
\be\label{EB}
E\geq E_B=\Omega_{(D)}\left(W(\phi(+\infty))-W(\phi(-\infty))\right),
\ee
where the $D$-dimensional solid angle contribution is given by $\Omega_{(D)} = 2\pi^{D/2}/\Gamma(D/2)$. The above procedure shows that the energy is minimized if the solutions obey the first-order equation
\begin{align}
\label{fo}
    \phi'=\frac{W_{\phi}}{r^{D-1}}+f.
\end{align}
We remark that, even though this formalism is only valid for monotonically increasing solutions, the first-order equations corresponding to the decreasing solutions can be found by taking $W(\phi)\to-W(\phi)$ and $f(\phi)\to-f(\phi)$ in Eqs.~\eqref{poten}--\eqref{fo}; this change does not modify the potential $V(\phi)$. Next, we investigate the stability of the model.

\section{Stability}\label{stabilitysec}
Since we are considering the inclusion of the cuscuton term in the Lagrangian density, we now turn our attention to stability of the solutions under small fluctuations. To do so, we take the field in the form $\phi(r,t)=\phi(r)+\eta(r,t)$, where $\phi(r)$ is the static solution and $\eta(r,t)$ is a time-dependent perturbation. The equation of motion \eqref{emtd} in this case takes the form
\be
\label{pertur}
  A^{-2}\Ddot{\eta}-\frac{1}{r^{D-1}}\big(r^{D-1}\eta'\big)'+\bigg(V_{\phi\phi}+\frac{D-1}{r}f_{\phi}\bigg)\eta=0,
\ee
where we have considered monotonically increasing solutions ($\phi'>0$) and used the notation $A^{-2}=(1-f/|\phi'|)$. We then proceed as usual and make the separation of variables $\eta(r,t)=\sum_{i}\eta_{i}(r)\cos(\omega_{i}t)$ to write the expression \eqref{pertur} as
\begin{align}
\label{stabilityeq}
      &-\frac{1}{r^{D-1}}\big(r^{D-1}\eta_{i}'\big)'\nonumber\\
      &+\bigg(V_{\phi\phi}+\frac{D-1}{r}f_{\phi}\bigg)\eta_{i}=A^{-2}\omega_{i}^2\eta_{i},
\end{align}
where
\begin{align}
    V_{\phi\phi}&=\bigg(f+\frac{W_{\phi}}{r^{D-1}}\bigg)
\bigg(f_{\phi\phi}\,+\frac{W_{\phi\phi\phi}}{r^{D-1}}\bigg) +\bigg(f_{\phi}+\frac{W_{\phi\phi}}{r^{D-1}}\bigg)^2.
\end{align}
The stability equation \eqref{stabilityeq} leads to stable solutions if $\omega^2_i\geq0$. It is a Sturm-Liouville  eigenvalue equation, so it can be written in the form $L\eta_{i}=\omega^{2}_{i}\eta_{i}$. In this case, the eigenfunctions obey
\be
\int^{\infty}_{-\infty}A^{-2}r^{D-1}dr \eta_i(r)\eta_j(r)=\delta_{ij}.
\ee
Notice that not only we have the factor $r^{D-1}$ in the above integral, but also the weight $A^{-2}$ which comes from the non-canonical character of our model \cite{kinkgen1,kinkgen2,stabkink}. One can show that $L=S^{\dagger}S$, where $S$ and $S^\dagger$ are adjoint operators given by
\bes
\bal
   S&=A\bigg[-\frac{d}{dr}+\bigg(\frac{W_{\phi\phi}}{r^{D-1}}+f_{\phi}\bigg)\bigg],\\
    S^{\dagger}&=A\bigg[\frac{d}{dr}+\bigg(\frac{W_{\phi\phi}}{r^{D-1}}+f_{\phi}\bigg)+\frac{\big(A^{-1}r^{D-1}\big)'}{A^{-1}r^{D-1}}\bigg].
\eal
\ees
Supposing that the above operators are smooth, the above factorization shows that only non-negative eigenvalues are present in Eq.~\eqref{stabilityeq}, so the solutions of Eq.~\eqref{fo} are stable against small fluctuations.

Since the results obtained in the above sections are general, let us now illustrate the procedure with some specific models.

\section{First model}\label{fm}
The first-order equation \eqref{fo} is quite intricate, as it engenders nonlinearity that may arise due to $W_\phi$ or $f$, and also the explicit dependence on the radial coordinate. Obtaining analytical solutions is, therefore, a hard task. Notwithstanding that, we have found an interesting class of models, with
\begin{align}
\label{fwphi}
    f(\phi)=\alpha\, W_{\phi},
\end{align}
where $\alpha$ is a positive-real parameter that controls the strength of the cuscuton term. The canonical model, in which the cuscuton term is absent, can be recovered for $\alpha=0$. In the more general case, however, we can write Eq.~\eqref{fo} as follows
\begin{align}
\label{fo111}
    \phi'= \left(\frac{1}{r^{D-1}}+\alpha\right)W_{\phi}.
\end{align}
We may change the variable of the above equation, defining
\be
\xi(r) = \int \frac{dr}{r^{D-1}}+\alpha r.
\ee
It makes Eq.~\eqref{fo111} become $d\phi/d\xi = W_\phi$, which is similar to the one that appears in the study of kinks in $(1,1)$ spacetime dimensions \cite{Vacha}. There, kinks arise as static solutions which connect neighbor minima of the potential. So, the presence of $\xi(r)$ allows us to map the $(1,1)$ kinks into the solutions $\phi(r)$ of Eq.~\eqref{fo111}. We remark that, in the above expression, an integration constant, which we call $r_0$, arises. To fix it, we shall take the condition $\xi(r_0)=0$. The integral in $\xi(r)$ depends on the number of spatial dimensions. For $D=2$, we have
\be\label{xi2}
\xi(r) = \ln\left(\cfrac{r}{r_0}\right) + \alpha (r-r_0).
\ee
For $D\geq3$,
\be\label{xi3}
\xi(r)=-\cfrac1{(D-2)}\left(\cfrac1{r^{D-2}}-\cfrac1{r_0^{D-2}}\right) + \alpha (r-r_0).
\ee
The last two expressions allow us to see that, regardless the number of spatial dimensions, $\xi(r)$ ranges from $-\infty$ to $+\infty$. This means that the $(1,1)$-dimensional kink can be fully mapped into our system, ensuring that the solution connects the minima of the potential. This cannot occur, however, in the case $\alpha=0$ with $D\geq3$, which was previously investigated in Ref.~\cite{prlbazeia}. There, the authors found that, in this situation, one has $\xi(r)\in(-\infty,0)$, showing that $\xi(r)$ only maps half of the $(1,1)$-dimensional kink.

The energy density \eqref{rhodens} can now be written as
\be\label{edens}
\begin{split}
	\rho &=\left(\left(\frac{d\xi}{dr}\right)^2  -\alpha \frac{d\xi}{dr}\right)\left(\frac{d\phi}{d\xi}\right)^2 \\
		&= \frac{1}{r^{D-1}}\left(\frac{1}{r^{D-1}}+\alpha\right)\left(\frac{d\phi}{d\xi}\right)^2
\end{split}
	\ee
We then consider henceforth the auxiliary function associated to the so-called $\phi^4$ potential
\be\label{wphi4}
W(\phi) = \lambda\phi-\frac{\lambda}{3}\phi^3.
\ee
This leads to $d\phi/d\xi=1-\phi^2$, whose associated solution is $\phi(r) = \tanh(\lambda\xi(r))$ and the associated energy density $\rho(r) = (\alpha + r^{1-D})r^{1-D}\sech^4\xi(r)$. We take $\alpha=f_0/\lambda$ to get how the cuscuton term really contributes in this solution. Next, we investigate the case $D=2$ and $D\geq3$ separately.

\begin{figure}[t!]
\centering
\includegraphics[width=6.5cm]{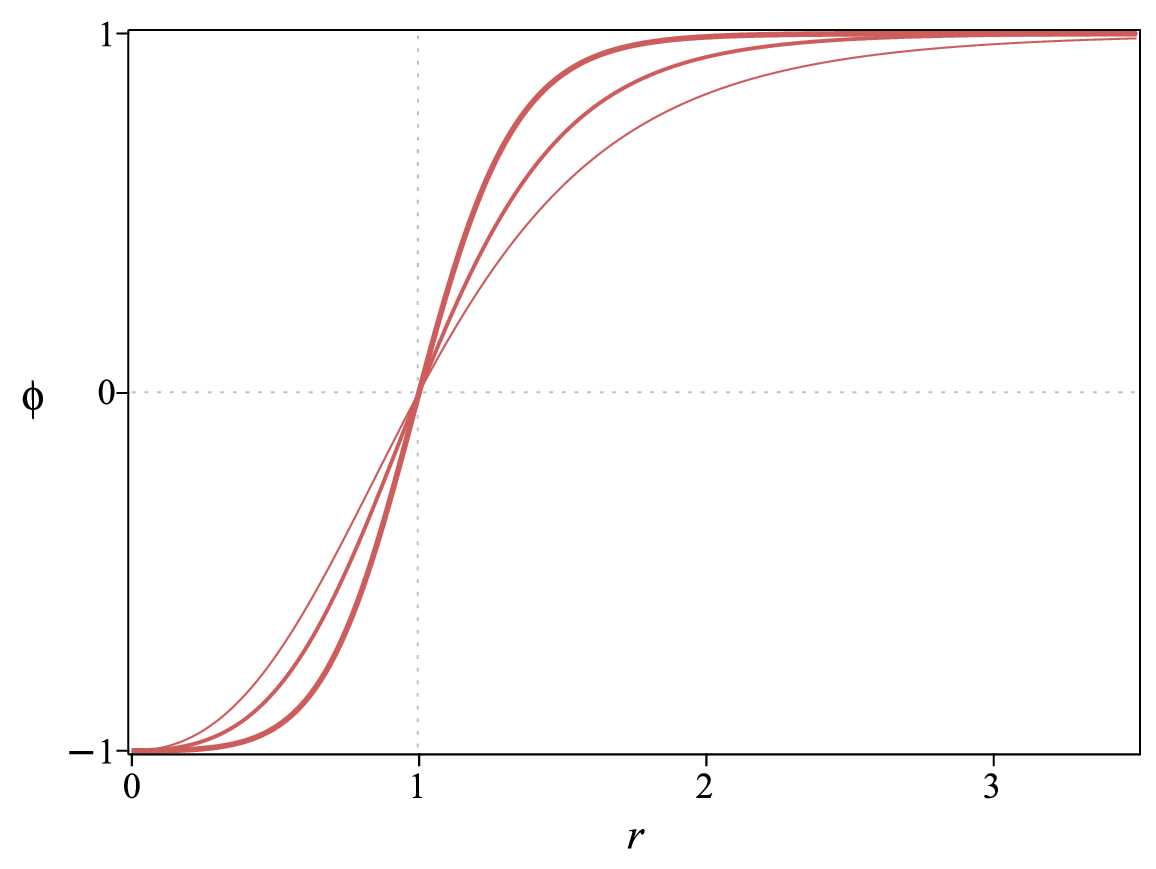}
\includegraphics[width=6.5cm]{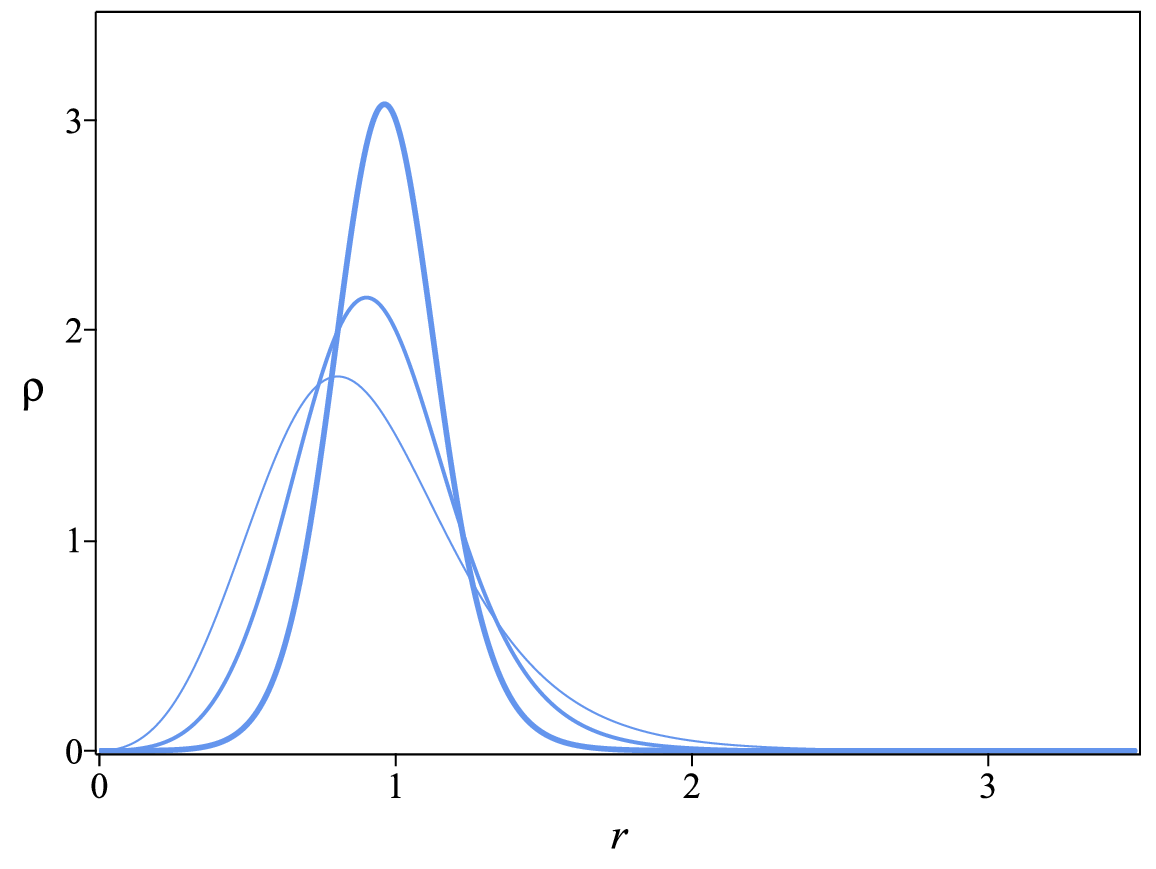}
\caption{The solution \eqref{solpag1} (top) and the energy density \eqref{rhopag1} (bottom) for $\lambda=r_0=1$ and $f_0=0.5,1$ and $2$. The thickness of the lines increases with $f_0$.}
\label{figpag1}
\end{figure}

In two spatial dimensions, the solution associated to the auxiliary function \eqref{wphi4} can be explicitly written as
\be\label{solpag1}
\phi(r) = \tanh\left(\lambda \ln\left(\frac{r}{r_0}\right) + f_0 (r-r_0)\right).
\ee
Therefore, $\phi$ ranges from $-1$ to $1$, which are the minima of the potential. Near the origin, the above expression behaves as
\be
\phi_{ori}(r)\approx-1+2\left(\frac{r}{r_0}\right)^{2\lambda} + {\cal O}\left[\left(\frac{r}{r_0}\right)^{4\lambda}\right].
\ee
Asymptotically, we have
\be\label{phiasy1}
\phi_{asy}(r)\approx 1-2e^{-2f_0(r-r_0)} + {\cal O}\left[e^{-4f_0(r-r_0)}\right].
\ee
This means that, even though the solution engenders a power-law profile near the origin, its asymptotic behavior is described by an exponential falloff. From the last expression, we can see that the tail of the solution depends only on the parameter associated to the cuscuton term, $f_0$. The energy density \eqref{edens} gets the form
\be\label{rhopag1}
\rho(r) = \frac{1}{r}\left(\frac{1}{r}+\frac{f_0}{\lambda}\right)\sech^4\left(\lambda \ln\left(\frac{r}{r_0}\right) + f_0 (r-r_0)\right).
\ee
In Fig.~\ref{figpag1}, we display the solution \eqref{solpag1} and the above energy density. We remark that the peak of the energy density gets taller and approaches $r=r_0$ as we increase $f_0$.

\begin{figure}[t!]
\centering
\includegraphics[width=6.5cm]{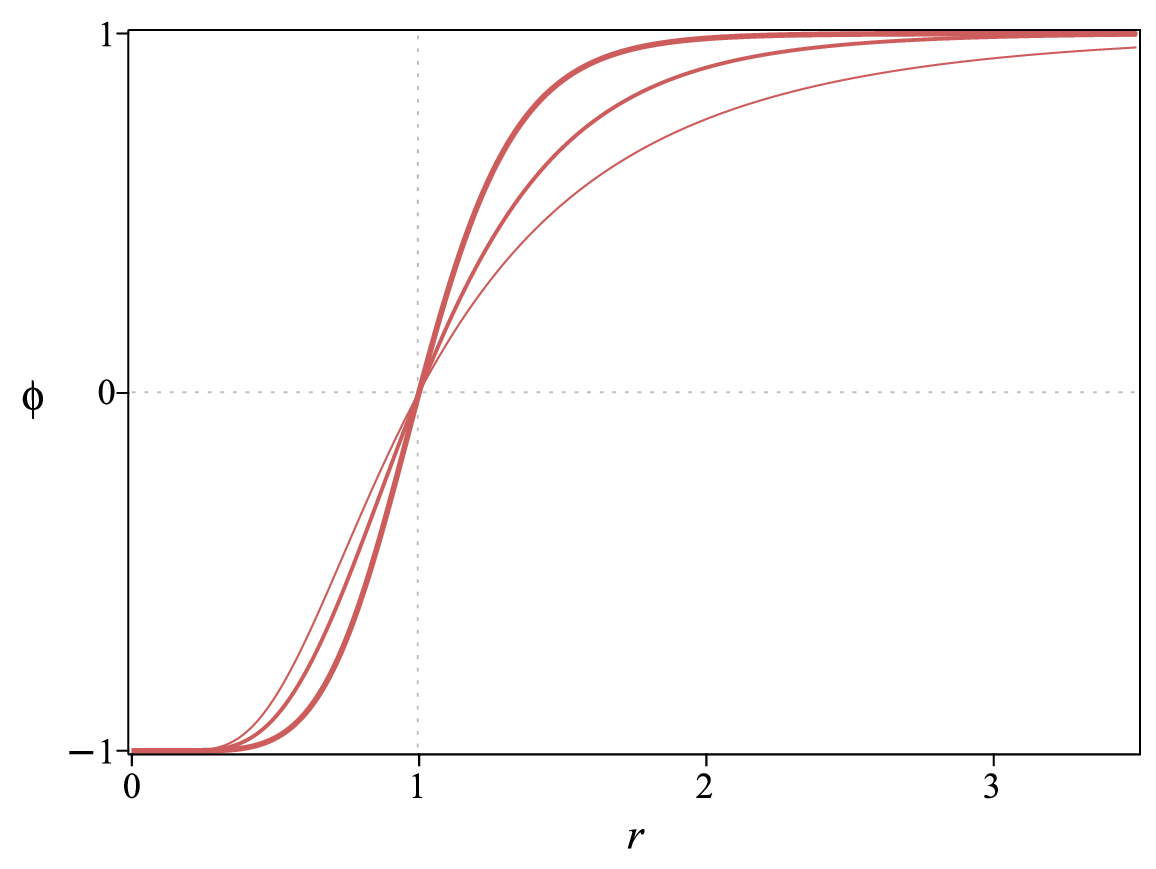}
\includegraphics[width=6.5cm]{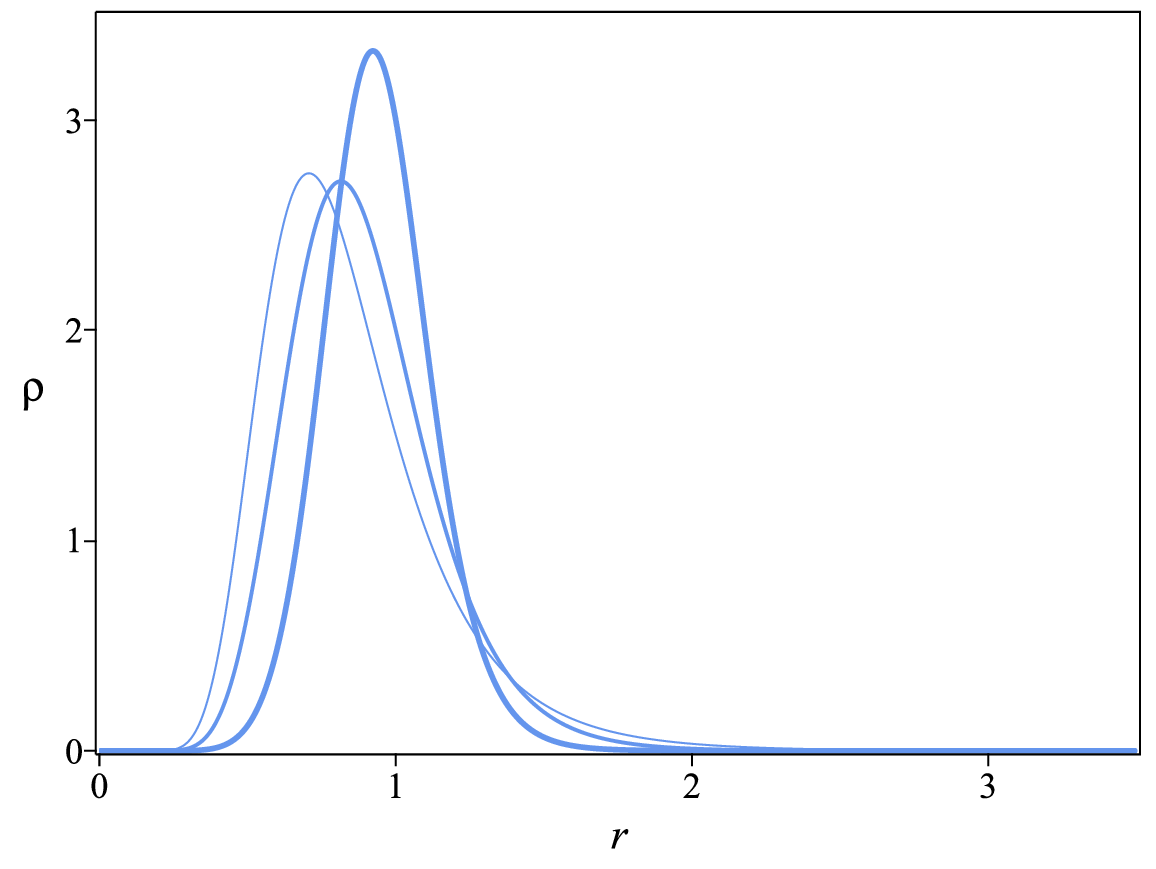}
\caption{The solution \eqref{solpag2} (top) and the energy density \eqref{rhopag2} (bottom) for $D=3$, $\lambda=r_0=1$ and $f_0=0.5,1$ and $2$. The thickness of the lines increases with $f_0$.}
\label{figpag2}
\end{figure}

We now turn our attention to the case $D\geq3$, for which we have the solution
\be\label{solpag2}
\phi(r) = \tanh\!\left(-\frac{\lambda}{(D-2)}\left(\frac{1}{r^{D-2}}-\frac{1}{r_0^{D-2}}\!\right) + f_0(r-r_0)\right),
\ee
where we have taken $\phi(r_0)=0$. Near the origin, one can show that it behaves as
\be
\phi_{ori}(r)\approx -1+2e^{-\frac{2\lambda}{D-2}\,\frac{1}{r^{D-2}}} + {\cal O}\left[e^{-\frac{4\lambda}{D-2}\,\frac{1}{r^{D-2}}} \right].
\ee
On the other hand, the asymptotic behavior is the very same of the one in Eq.~\eqref{phiasy1}. This is due to the fact that the term which arises in $\xi$ due to the cuscuton term does not depend on the number of dimensions; see Eqs.~\eqref{xi2} and \eqref{xi3}. The energy density \eqref{edens} reads
\be\label{rhopag2}
\begin{aligned}
	\rho(r) &= \frac{1}{r^{D-1}}\left(\frac{1}{r^{D-1}}+\frac{f_0}{\lambda}\right)\\
	&\times\sech^4\!\left(-\frac{\lambda}{(D-2)}\!\left(\!\frac{1}{r^{D-2}}-\frac{1}{r_0^{D-2}}\!\right) + f_0(r-r_0)\!\right).
\end{aligned}
\ee
In Fig.~\ref{figpag2}, we display the solution \eqref{solpag2} and \eqref{rhopag2} in three spatial dimensions. We emphasize that, contrary to the case $D=2$, we cannot take $f_0=0$ for $D\geq3$ because it would lead to a solution that does not connect the minima of the potential. In the interval $f_0>0.78$, as $f_0$ gets larger and larger, i.e., in the regime in which the cuscuton term is more and more relevant, the energy density tends to concentrate around $r=1$ and its peak becomes taller.

\section{Second model}\label{sm}
Obtaining a solution, as in the previous section, for a general $f(\phi)$ is quite hard due to the explicit presence of the coordinate in the first-order equation \eqref{fo}. In this section, we propose models in two and higher spatial dimensions that escape from the choice in Eq.~\eqref{fwphi}.

We start considering the case $D=2$ with a power-law solution in the form
\be\label{solmatrob2}
\phi(r) = \frac{r^{2\beta}-r_0^{2\beta}}{r^{2\beta}+r_0^{2\beta}}.
\ee
Here, $\beta$ is a real parameter and $r_0$ determines the point in which the solution reaches zero, that is, $\phi(r_0)=0$. This solution ranges from $\phi(0)=-1$ to $\phi(\infty)=1$.
The behavior at the origin is
\be
\phi_{ori}(r)\approx-1+2\left(\frac{r}{r_0}\right)^{2\beta} + {\cal O}\left[\left(\frac{r}{r_0}\right)^{4\beta}\right].
\ee
The asymptotic behavior is
\be
\phi_{asy}(r)\approx1-2\left(\frac{r_0}{r}\right)^{2\beta}  +{\cal O}\left[\left(\frac{r_0}{r}\right)^{4\beta}\right]
\ee
The latter two expressions show us that, in order to avoid divergences in the first and second derivatives of the solution, one must impose $\beta\geq1$.

To construct a model that supports the solution \eqref{solmatrob2}, we continue to use the auxiliary function as in the form Eq.~\eqref{wphi4}. We have found that this solution is compatible with the first-order equation \eqref{fo} for
\be\label{fcusc2}
f(\phi) = f_0\left(1-\phi\right)^{1+\frac{1}{2\beta}}(1+\phi)^{1-\frac{1}{2\beta}},
\ee
where $r_0=(\beta-\lambda)/f_0$, with $\beta>\lambda$ to ensure $r_0>0$. Notice that the above function vanishes at the boundary values of the solution \eqref{solmatrob2}.

\begin{figure}[t!]
\centering
\includegraphics[width=6.6cm]{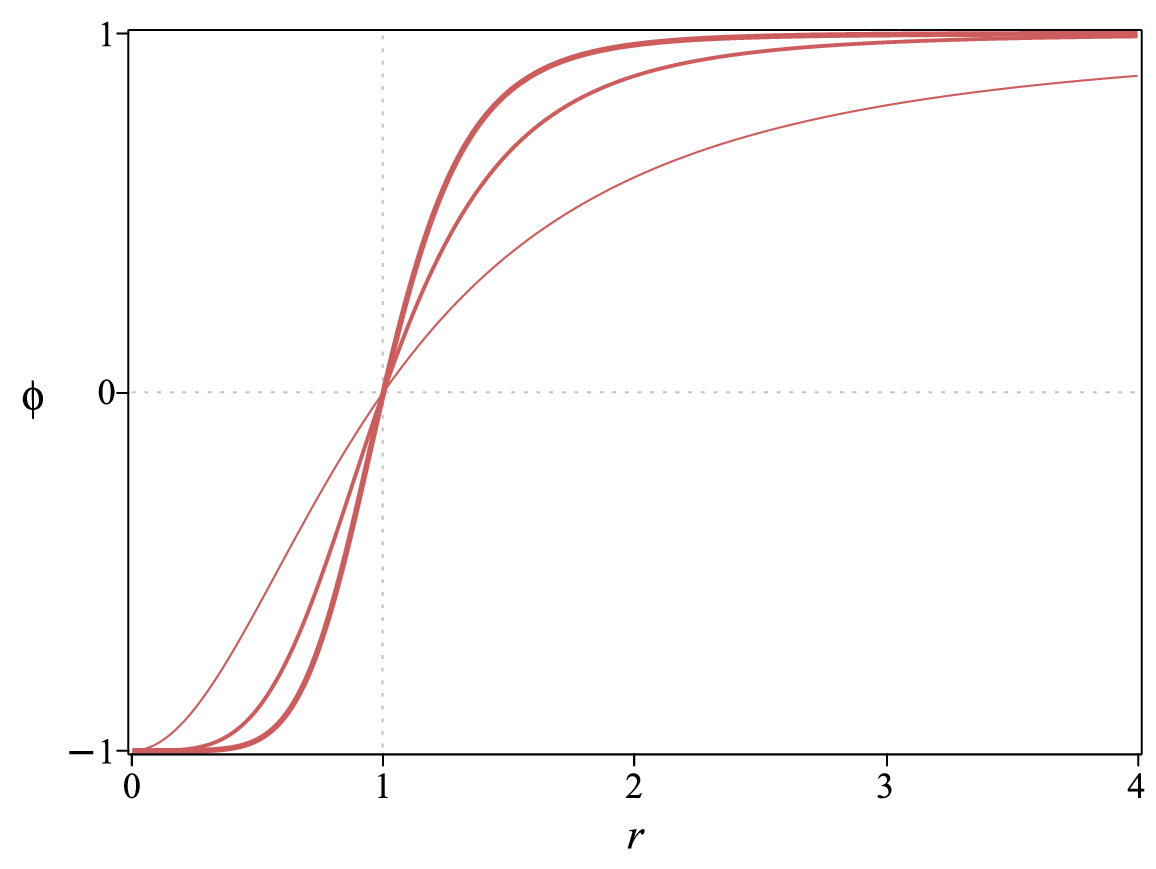}
\includegraphics[width=6.5cm]{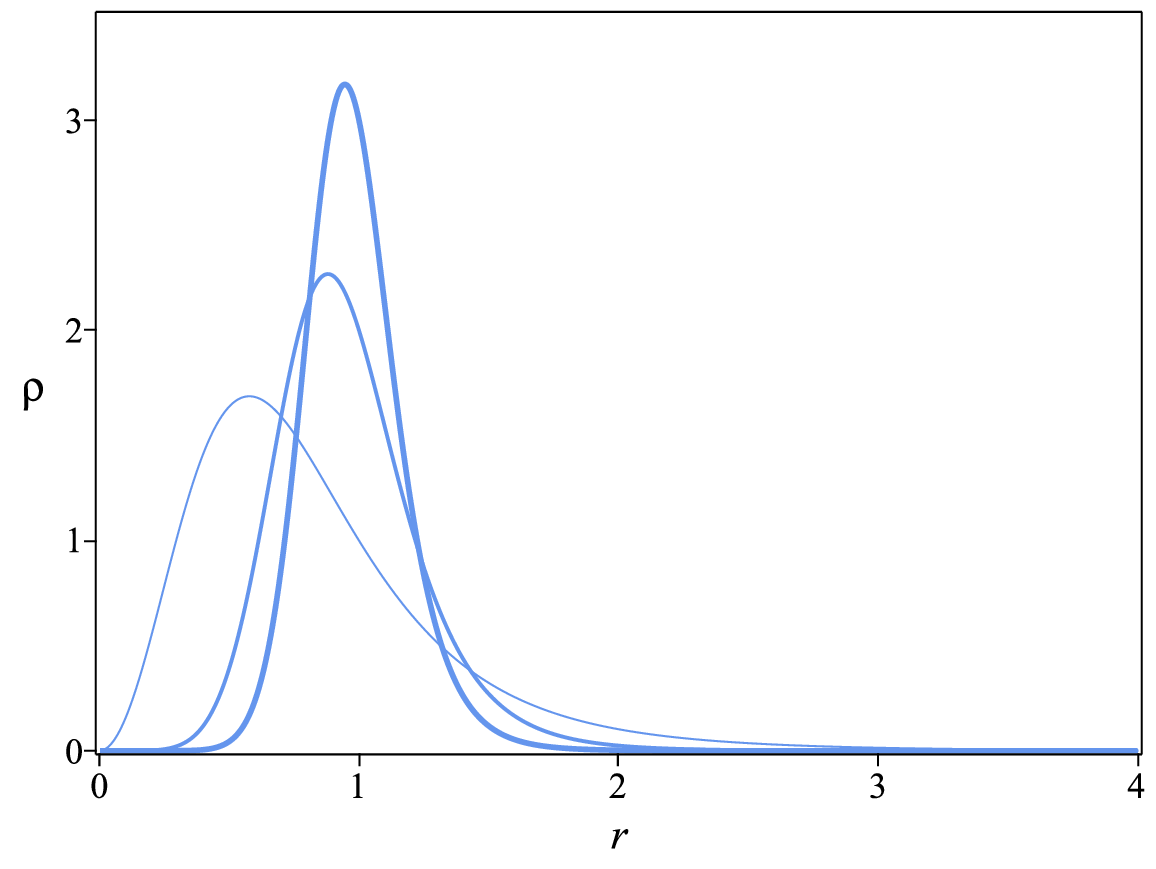}
\caption{The solution \eqref{solmatrob2} (top) and the energy density \eqref{rhomatrob} (bottom) for $\lambda=r_0=1$ and $\beta=1,2$ and $3$. The thickness of the lines increases with $\beta$.}
\label{figmr}
\end{figure}

To consider the solution \eqref{solmatrob2} in $D\geq2$ with the same cuscuton term, i.e., the same function in Eq.~\eqref{fcusc2}, we must change the auxiliary function to
\be
\begin{aligned}
	W(\phi) &= \frac{2^{2+\frac{D-2}{2\beta}}\lambda\beta}{D-2-4\beta}\left(1-\phi\right)^{2-\frac{D-2}{2\beta}}\times\\
	& {}_2F_1\!\left(2-\frac{D-2}{2\beta},-1-\frac{D-2}{2\beta};3- \frac{D-2}{2\beta};\frac12-\frac{\phi}{2}\right)\!,
\end{aligned}
\ee
where ${}_2F_1(a,b;c;z)$ denotes the hypergeometric function. By taking the derivative of the above expression, one gets
\be
W_\phi= \lambda \left(1+\phi\right)^{1+\frac{D-2}{2\beta}}(1-\phi)^{1-\frac{D-2}{2\beta}},
\ee
From which we see that the parameter $\beta$ must obey $\beta>(D-2)/2$ to avoid divergencies. In this model, we must impose that $r_0$ is the solution of 
\be\label{constpar}
f_0r_0+\lambda r_0^{D-2}=\beta.
\ee
In this case, the energy density is
\be\label{rhomatrob}
\begin{split}
	\rho(r) &= \lambda\beta r_0^{-D}\left(1-\phi(r)\right)^{2+\frac{1}{\beta}}(1+\phi(r))^{2-\frac{1}{\beta}}\\
	&=\frac{16\lambda\beta r_0^{4\beta+2-D}r^{4\beta-2}}{\left(r^{2\beta}+r_0^{2\beta}\right)^4}.
\end{split}
\ee
Notice that the signature of the number of spatial dimensions only appears in the power of $r_0$. The energy is 
\be
\begin{aligned}
	E &= -\frac{(D-2)(D-2+2\beta)(D-2(1+\beta))\pi}{6\beta^3}\\
	&\times\csc\left(\frac{(D-2)}{2\beta}\,\pi\right) 
\end{aligned}
\ee
The energy for the case $D=2$ can be obtained by taking the limit $D\to2$ in the above expression, which leads us to $E=4/3$. In Fig.~\ref{figmr}, we display the solution \eqref{solmatrob2} and the energy density \eqref{rhomatrob} for $r_0=\lambda=1$ and some values of $\beta$. We remark that plots depicted in this figure are valid for any number of spatial dimensions because the choice $r_0=1$ makes $D$ disappear in the expressions. Also, since the parameters must obey Eq.~\eqref{constpar}, we have $f_0=0,1$ and $2$ for $\beta=1,2$ and $3$, respectively, when $r_0=\lambda=1$.

\section{Ending Comments}\label{finalremarks}
In this work, we have investigated scalar field models in the presence of the cuscuton term in $(D,1)$ spacetime dimensions. By considering that scalar field engenders radial symmetry, we have found the equations of motion and the associated energy density for static configurations. The results show that, even though the cuscuton term does not contribute in a single spatial dimension, one must take into account an extra term in the equation of motion for higher dimensions. To simplify the problem, we have developed a first-order formalism based on the energy minimization which leads to the energy depending only on the auxiliary function calculated at the boundary values of the field. Since the model under investigation has a non-canonical character, we have also studied the stability of the static solutions around small fluctuations. In this case, one gets a Sturm-Liouville eigenvalue equation. We have then factorized the operator of the aforementioned equation in terms of adjoint operators to show that the general model is linearly stable due to the absence of negative eigenvalues. This is an interesting result, indicating that the presence of the cuscuton term does not necessarily destroy stability of the solutions in arbitrary dimensions. 

To illustrate the procedure, we have investigated a class of models obeying Eq.~\eqref{fwphi}. The first-order equation that drives the model can be mapped into the one which describes kinks in $(1,1)$ spacetime dimensions in canonical models. Interestingly, the cuscuton term contributes to make the full mapping of the solutions. When it is absent, $f(\phi)=0$, the case $D\geq3$ can only be mapped in one half of a kink. By considering the auxiliary function associated to the so-called $\phi^4$ model, we have shown that the asymptotic behavior of the solutions does not depend on $D$, as it is controlled exclusively by the parameter related to the cuscuton term. Next, we have considered a second class of models, in which we show the conditions to get a solution with power-law tails. In this case, the parameters are constrained and the auxiliary function depends on the number of dimensions. The results for both the solutions and energy densities are all analytical, and they are displayed in the three figures that appear in the work.

As perspectives, since the cuscuton term leads to scalar field solutions connecting the minima of the $\phi^4$ potential, one may consider to take it in the study of global monopoles \cite{global1,global2}, where solutions may not connect the minima of the potential. Another interesting possibility is to investigate electrically charged structures in the presence of the cuscuton term \cite{Bazeia:2022pon} and new ways of coupling the scalar field to the gauge field. Moreover, one can study its influence on the scenario of models that includes the Higgs portal in the context of topological structures \cite{Arcadi:2019lka}. One may also consider the presence of the cuscuton term in Bose-Einstein Condensates, to verify how it modifies the solitons in these systems, where the density and dispersion play an important role in the stability of the configurations \cite{bec,bec2}. We are now considering some of the above possibilities, hoping to describe novel results in the near future.

\acknowledgements{This work is supported by the Brazilian agencies Conselho Nacional de Desenvolvimento Cient\'ifico e Tecnol\'ogico (CNPq), grants 402830/2023-7 (DB, MAM, RM), 303469/2019-6 (DB), 306151/2022-7 (MAM) and 310994/2021-7 (RM).}

\end{document}